# Advanced Halide Scintillators: From the Bulk to Nano

*Vojtěch Vaněček,\* Kateřina Děcká, Eva Mihóková, Václav Čuba, Robert Král, and Martin Nikl\**

**Halide scintillators have been playing a crucial role in the detection of ionizing radiation since the discovery of scintillation in NaI:Tl in 1948. The discovery of NaI:Tl motivated the research and development (R&D) of halide scintillators, resulting in the development of CsI:Tl, CsI:Na, $CaF_2$:Eu, etc. Later, the R&D shifted toward oxide materials due to their high mechanical and chemical stability, good scintillation properties, and relative ease of bulk single-crystal growth. However, the development in crystal growth technology allows for the growth of high-quality single crystals of hygroscopic and mechanically fragile materials including $SrI_2$ and $LaBr_3$. Scintillators based on these materials exhibit excellent performance and push the limits of inorganic scintillators. These results motivate intense research of a large variety of halide-based scintillators. Moreover, materials based on lead halide perovskites find applications in the fields of photovoltaics, solid-state lighting, and lasers. The first studies show also the significant potential of lead halide perovskites as ultrafast scintillators in the form of nanocrystals. The purpose of this review is to summarize the R&D in the field of halide scintillators during the last decade and highlight perspectives for future development.**

## 1. Introduction

Research and development (R&D) of solid-state scintillation materials has been an intense research topic in the past three decades. It was launched by the needs of high-energy physics at the beginning of 1990s which needed radiation-resistant materials for the new generation of accelerators (Large Hadron Collider in CERN) and further intensified by the search for heavy, fast, and efficient scintillators to be applied in medical imaging (CT, positron emission tomography [PET], SPECT) and most recently by needs of homeland security techniques to counterfight illicit traffic of radioisotope materials, explosives, and drugs, and to prevent the population from terrorist attacks. Last but not least, there are specific fields in the industry (defectoscopy, particle beam detection and diagnostics, geophysical and environmental applications, to name a few), the progress of which is also based on the R&D of tailored scintillators for each application. Such a broad variety of applications results in rich and specific demands on scintillation parameters in tailored material structures which explain a long-lasting R&D coming all the time with new issues to solve.

Scintillation material works as a transformer of high-energy photons, accelerated particles, or even neutrons into a bunch of UV/VIS/NIR photons (a flash of light), **Figure 1**, which are easily detectable by existing photodetectors (photomultipliers, semiconductor-based devices). The scintillation mechanism, that is, the physical scheme behind this transformation process, consists of three consecutive stages, namely, the conversion, transfer, and luminescence, described in more detail in numerous existing reviews which are also focused on various application fields, material groups, or specific parameters.[1–6]

In the 1990s, R&D was focused mainly on oxide-based materials due to the development of $PbWO_4$ single crystals for high-energy physics and Ce-doped orthosilicates and perovskites for medical imaging. Around the beginning of the new millennium, attention has also been paid to new families of halide scintillators based on rare earth (RE) halides, see the study by Kramer et al.,[7] followed by the search for efficient and high-energy-resolution materials in $Eu^{2+}$-doped bromides and iodides,[6] high-$Z_{eff}$ halide materials, the composition of which includes Hf[8] and Tl[9] elements, and the most recent extension of classical CsI scintillator toward ternary Cs–Cu–I compositions[10] with advanced characteristics. A survey of scintillation parameters of the bulk halide single crystals is summarized in Table S1, Supporting Information.

In addition to the bulk single-crystal form of these materials, microstructured halide materials in eutectic form with light-guiding property, nanocrystals (NCs), nanostructures, and nanocomposites were reported, which further enriched the halide scintillator material family. Probably, the most intense R&D in recent years is devoted to halide perovskites including especially the nanomorphological materials. Luminescence of

V. Vaněček, K. Děcká, E. Mihóková, R. Král, M. Nikl
FZU Institute of Physics
Czech Academy of Sciences
Cukrovarnicka 10, 16200 Prague, Czech Republic
E-mail: vanecekv@fzu.cz; nikl@fzu.cz

V. Vaněček, K. Děcká, E. Mihóková, V. Čuba
FNSPE
Czech Technical University in Prague
Brehova 7, 11000 Prague, Czech Republic

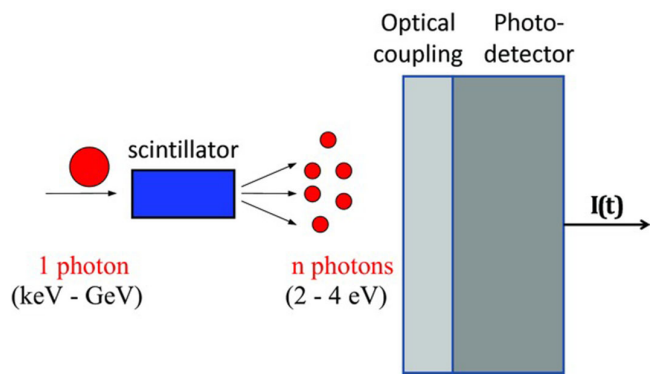

**Figure 1.** A sketch of the scintillation detector. Reproduced with permission.[5] Copyright 2021, John Wiley and Sons.

classical halide perovskites, $CsPbX_3$ (X=Cl, Br) single crystals prepared by the Bridgman method, was studied a few decades ago including the luminescence decay kinetics.[11,12] Even the luminescence of nanosized quantum dots (QDs) of these compounds was reported in the 1990s, spontaneously created after thermal annealing of Pb-doped CsX (X=Cl, Br) single crystals.[13,14] Their intrinsic emission is due to the Wannier exciton characterized by subnanosecond lifetimes and small Stokes shift of several tens of meV, which disables guiding the luminescence light out of bulk elements. Recently, a family of halide perovskites has emerged as an interesting class of semiconductors, highly promising for a variety of applications. Single crystals and polycrystalline thin films have been extensively investigated in photovoltaic cells and photodetectors thanks to their strong light absorption, high charge carrier mobility, and exceptional defect tolerance. However, resulting from mobile ionic defects and small exciton binding energy, their PL quantum yields (PLQYs) at room temperature (RT) are low, limiting thus their light-emitting applications.[15,16]

The way to significantly improve PLQY appears in the preparation of halide perovskites in the form of self-standing NCs (NCs), having higher exciton binding energy and lesser defects.

Nanostructured perovskite materials are of great interest due to multiple quantum-confinement effects. The most widely known perovskite NCs are the perovskite QDs which have sizes in the nanometer scale that are less than or close to the exciton Bohr radius. In terms of emission, perovskite QDs have a higher color purity, a wider color gamut, and lower cost of processing compared with traditional QDs.

Starting from the first reports on colloidal solutions of organic–inorganic (hybrid) perovskite NCs[17] and especially fully inorganic perovskite NCs,[18] immense research efforts have been devoted to the study and development of these materials. Due to their excellent luminescence properties, they became instantly identified as highly promising for a number of applications, such as solar cells,[19–23] photodetectors,[24–26] light-emitting diode (LEDs),[27–32] displays,[33–36] and lasers.[37–39] Recently, a body of studies on lead halide perovskites has also been focused on their application in scintillation detectors; see recent review papers.[40–43]

The aim of this review is to provide a broad overview in the field of halide scintillators studied in the past decade focusing on material composition and its scintillation performance and also on new trends in nonsingle-crystal systems' R&D trying to explain the added value which they might bring if their manufacturing technology will advance enough.

## 2. Bulk Crystal Growth of Halide Materials

An important part of the R&D of halide scintillators is their preparation in the form of single crystals. Throughout history, various technologies and methods for crystal growth of halide materials have been developed. However, the hygroscopic nature of halides complicates the manufacturing process contributing to the formation of various oxidic impurities and thus influencing the behavior of halide's melt (e.g., wetting the growth container, cracking of crystal or the container) and degrading the final optical and physical properties of prepared scintillators. Generally, the hygroscopicity of halide salts is an ability to absorb moisture from ambient atmosphere,[44] depending on the temperature, relative humidity, and time of exposure.[45] The hygroscopicity of halide salts can be roughly estimated by the salt's solubility in water.[46]

The origin of oxidic impurities comes from reactions of halides with ambient atmosphere, moisture, and oxygen under the formation of, for example, hydrates, carbonates, hydroxy- and oxy-halides, and oxides.[44,46–49] Removal of such impurities is absolutely essential before the growth to prepare high-quality and pure single crystals. Thus, multiple techniques were developed to achieve such a goal consisting for instance of introduction of halogenation agents into molten halides,[46,48–50] implementation of vacuum to the raw material in a container under thermal treatment,[51–53] application of zone refining,[54–56] or their combination.[47,50]

Interestingly, the technology of introducing halogenation agents into molten halides was independently developed and presented in similar period in 1960s' and 1970s' by Lebl and Trnka,[48] Rosenberger,[46] and Kanzaki and Sakuragi,.[57] Generally, the process is based on the multiple step reactions of various reagents such as hydrogen halides (HCl, HBr),[46,48,56] halogens ($Cl_2$, $Br_2$, $I_2$),[46,48] organic halides (e.g., $COCl_2$, $CCl_4$, $CHCl_3$),[48] silicon tetrachloride,[49] and others, with oxidic impurities in molten halides and their replacements with the required halide anion. Other purification techniques may be used as well such as 1) filtration of solutions as a basic removal of insoluble particles; 2) extraction of certain constituents based on preferential dissolution in a solvent; 3) liquid–liquid extraction based on the distribution of a solute (i.e., impurity in a dissolved matrix) between two essentially immiscible solvents; 4) ion exchange by running a solution through an inert tube filled with a resin enabling cationic separation in alkali halide solutions, as well as 5) precipitation of insoluble compounds from certain constituents in a solution by introduction of a selectively reacting agent. For more details, see the study by Rosenberger.[46]

The most widely used technique for bulk crystal growth of halides from their melt is the vertical Bridgman (VB) method. This technique was developed by Bridgman and published in 1925[58] belonging into a group of directional solidification methods. It is based on placing a container (crucible or ampoule) with the raw material into a tubular vertical furnace heated above the melting point of the material. Afterward, the container is pulled down out of the furnace, where solidification appears. Further, a very important modification was suggested by Stockbarger[59] for the growth of LiF using two independently heated tubular and vertically arranged furnaces separated with an annular baffle. The upper furnace (or a so-called zone) was heated above the material's melting point and the lower one below creating a well-defined temperature gradient between them. The growth itself proceeded similarly by lowering the container with molten LiF from the upper furnace through the gradient into the lower furnace. Such a technique with minor changes (e.g., number of furnaces/zones, heating elements, etc.) is used up to now and it is most commonly referred to as the VB–Stockbarger method or simply the Bridgman method. Recently, the design of the technique was adjusted for simultaneous multiple-ampoule growth of $KCaI_3$ using the so-called multiampoule growth station (MAGS).[60,61] Such multiampoule furnace designs incorporating an inductively or resistively heated crucible for growth of GaAs as well as fluoride-based materials have been proposed for instance in other studies.[62,63] Importantly, the Bridgman technique enables to select the type of the container used for the growth procedure. Depending on the properties of the grown material (such as melting point, reactivity, hygroscopicity, etc.), the container can be either opened, for example, alumina, platinum, graphite crucible, or closed, for example, sealed quartz ampoule. However, only the latter one is perfectly suitable for growth of hygroscopic halides as it prevents any contact with ambient atmosphere. In this way, various single crystals of halide scintillators were prepared, for instance, NaI:Tl,[64] $SrI_2$:Eu,[51,65,66] $LaBr_3$:Ce, and[67,68] various ternary and mixed halides and eutectics: $^6LiCl/^6Li_2SrCl_4$,[69] $Tl_2LaCl_5$:Ce,[70] $KSr_2I_5$:Eu,[71] $Cs_2HfCl_6$,[8,72,73] $Cs_3Cu_2I_5$,[53,74] $Ba_2CsI_5$:Eu,[75] BaBrI:Eu.[76]

The Czochralski (CZ) method is another crystal growth technique widely used in the R&D and commercial production of large-sized single crystals. It was developed by J. CZ in 1918[77] and since then it was applied for the production of various single crystals such as semiconductors (Si, Ge),[78] sapphire,[79] multicomponent oxides (YAG:Ce, GGAG:Ce),[80,81] and many others. Originally, the method was invented for measuring the crystallization velocity of metals by preparing them in the form of single-crystalline wires.[77,82] However, significant improvements based on the CZ method to produce controlled bulk semiconductor single crystals of germanium were proposed by Teal and Little.[83,84] Since then, the method remained unchanged with minor modifications until today. In summary, the CZ technique consists of placing a starting charge into a crucible (e.g., metals, ceramics, graphite, etc.), heating it above its melting point, tipping the melt with a seed (e.g., parent material, metal wire, etc.) from above, and slowly pulling the seed up again and rotating it, while crystallization takes place and the new crystalline phase is grown at the melt's surface. Even if this method was discovered about 100 years ago, its application for the growth of halide crystals was introduced only recently. Apart from using an air-tight CZ system, the procedure requires additional modifications due to hygroscopicity of halides such as baking of the growth chamber prior to the experiment or treating it by a high-purity carrier gas (e.g., Ar) through a molecular sieve; for more details, see, e.g., an article related to the preparation of mixed halides BaBrCl and BaBrCl:Eu.[85] A similar procedure was applied in the case of $PbI_2$,[86] eutectics CsI/CsCl/NaCl,[87] or halides with higher hygroscopicity such as $CsCe_2Cl_7$, $Cs_2NaCeCl_6$,[88] and $SrI_2$:Eu.[89]

The Kyropoulos method is a growth technique derived from the CZ technique and is up to now frequently used for industrial production of large-sized single crystals. It was developed by Kyropoulos[90] and it is based on a similar arrangement as in the CZ method, that is, a crucible with starting charge is heated above its melting point, seed touched at its surface from above, and afterward the crystal growth is commenced. However, the growth is conducted with following differences from the CZ method[91]: 1) complete crystallization of the melt in the crucible is carried out; 2) the final shape of the crystal is defined by the shape of the crucible; 3) the seed is water cooled; 4) rotation of the seed takes part only at the beginning of the growth; 5) a slow pulling of the seed up is performed only at the initial part of the growth followed by controlled cooling of the system; and 6) pulling is stopped when the crystal reaches bottom of the crucible. In this way, the single crystals of various sizes (small to large) of semiconductors, for example, ZnTe,[92] sapphire,[93] including halides such as CsBr (9 inch diam.),[94] KX:Cu (X=Cl, Br, I),[95] or LiF,[96] were prepared.

The edge-defined film-fed growth or simply edge-defined film-fed growth (EFG) method is a technique developed by H. LaBelle[97] for the growth of sapphire fibers used as reinforcements in metal matrix composites. This technique is similar to the CZ method as it is based on melting the starting charge in a crucible and placing a floating capillary die (made of e.g., refractory metal, carbon, etc.) with multiple channels on the melt's surface. Then, the melt is brought to the top of the die by capillary forces, where it is often spread over the die. Afterward, a seed touch is performed and the seed is pulled up. The crystal/melt interface is stabilized at the die and fed by a thin film of the melt coming out of the capillaries. Due to the melt's wetting of the die toward its edge, the final shape of the crystal is defined by the shape of the die.[91] Thus, EFG is a suitable and widely used technique for the production of shaped crystals (e.g., tubes, rods – circular or rectangular, sheets, etc.) of various materials (dielectric, semiconductors, etc.) of different sizes. Nevertheless, the origin of a shaped crystal growth can be tracked even to earlier times to Stepanov,[98] who suggested a similar technological approach. The differences between EFG and Stepanov method, as mentioned by LaBelle,[99] can be simplified to that the EFG process must use a wettable die and the Stepanov technique generally uses a nonwetted shaper. This is not entirely true as the key to Stepanov's technique is shaping a melt column in the die, which also defines the final shape of the grown crystal. On the contrary, in EFG the shape of the die controls the shape of the crystal.[99] Apart from the growth of sapphire fibers,[97,99] and for instance of $Mn_2SiO_4$,[100] EFG was used for the growth of halide materials such as mixed ternary $CsSr_{1-x}Eu_xI_3$,[101] eutectics (e.g., NaCl–LiF, LiF–NaF),[102] $SrI_2$,[103,104] TlBr,[105] or CsI.[106,107]

The micropulling-down method (μ-PD) is a modern technique proposed by Ohnishi et al.[108] for fiber crystal growth of $LiNbO_3$. Afterward, the technique was adjusted for the growth of different materials such as mixed Si–Ge,[109] $PbWO_4$,[110] multicomponent garnets,[111,112] LiF,[113] or metals and alloys.[114] In the case of hygroscopic halides, an important modification of μ-PD was reported by Yokota et al.[115] for the growth of NaCl:Ce,Pr.

The method was modified and equipped with a removable growth chamber enabling its transport from the apparatus into an oxygen and moisture-free atmosphere-controlled glovebox, where a setup of a hotzone, weighing of starting materials, and other handling procedures were performed. The growth itself was performed after the transport of the chamber back to the μ-PD apparatus in a similar way as for other materials mentioned earlier. The starting charge, placed in a crucible with a hole in its bottom, was heated above its melting point, touched with a seed from underneath (e.g., Pt wire), see **Figure 2**a, and pulled down below the crucible, where crystallization took place. Since then μ-PD was used for growth of, for example, $SrI_2$:Eu,[116] $CeCl_3$, and $LaBr_3$:Ce,[117] or ternary $RbPb_2Cl_5$:RE[118]. Due to high pulling rates (0.03–1.0 mm min$^{-1}$), low starting charges (up to 1 g), short experimental time (1–3 days), the μ-PD technique is ideal for an initial R&D and materials screening as it is significantly more cost effective than traditional growth techniques (VB, CZ, etc.). Recently, reports about modifications of μ-PD to grow halide crystals via the VB method of, for example, Eu:$SrI_2$[119] or $Cs_2HfCl_6$[120,121] were presented. This is the so-called miniaturized VB method (mVB), see Figure 2b. Such a setup combines advantages of both μ-PD and Bridgman methods, that is, fast experiment and cost effectivity and closed container using sealed ampoules, respectively.

Vertical gradient freeze method (VFG) is derived from the Bridgman technique by assembling a dynamically controlled multiple-zone furnace. The crystal growth is then performed by moving a thermal gradient through the container with the material instead of any mechanical translation mechanism of the container or the crystal like in the traditional methods (e.g., VB or CZ).[122] Such a setup minimizes thermal stress generation of dislocations using low thermal gradients, low growth rates, and controlling postsolidification cooling rates below the yield point for the stress-induced dislocation generation.[123] Initially, this technique was used for crystal growth of semiconductors and many publications were devoted to CdTe applying both the vertical[124] and horizontal[125] arrangements. On the contrary, the reports on the vertical gradient freeze (VGF) growth of halide crystals were limited to few publications regarding preparation of $KGd_2Cl_7$:Ce,[126] $Cs_3EuI_5$,[127] and CsI:Tl.[128]

The skull method is another technique belonging to the directional solidification ones developed in 1960s.[129–131] It is based on melting a material with inductive heating (contact-free) and keeping it in a cooled solid shell (skull) from the parent material, that is, with a chemical composition identical to that of the melt. Thus, such a method provides following advantages: no limit on the heating temperature (up to 3000 °C and higher), variability of the growth atmosphere (including an oxidizing one), growth of large-sized crystals, crystallization conducted by moving an inductive heat source or gradual reduction of power, and waste-free method (possibility of remelting unused scraps), for more details see the study by Osiko.[130] The method was widely used for purification and growth of refractory semiconductors and metals (e.g., Ge, Si, Zr),[132,133] growth of multicomponent oxides $(ZrO_2)_{0.89}(Sc_2O_3)_{0.1}(CeO_2)_{0.01}$ crystals,[134] as well as halide crystals such as large-sized NaI:Tl.[135,136]

## 3. Divalent Rare-Earth-Activated High LY and ER Halide Scintillators

In the last decade, there was an intense R&D of the $Eu^{2+}$-activated scintillators to discover a scintillator with an ultimate light yield (LY) and energy resolution (ER) mainly for the nuclear nonproliferation applications. Fewer reports also deal with the $Yb^{2+}$- and $Sm^{2+}$-doped halides. All these centers' luminescence is based on the $5d$–$4f$ transitions. In the case of $Yb^{2+}$, given its electronic configuration in the $4f$ shell, the spin-allowed (low spin: LS) and spin-forbidden (high spin: HS) transitions can be resolved in the excitation and emission spectra in which the HS-related ones are always low energy shifted compared with LS ones; see the study by Dorenbos et al.[137] The lifetimes of HS-based emissions are of the order of tenths of ms, while those related to LS ones are about two orders of magnitude faster; see systematic reports made on the $Yb^{2+}$-doped $CsCaX_3$ and $CsSrX_3$ (X = Cl, Br, and I).[138–140] The decay times of LS-based transitions are strongly temperature dependent and appear as a result of a delicate interplay of symmetry and also nonradiative relaxation pathways interconnecting the excited HS and ground states of the $Yb^{2+}$ center. In the case of $Sm^{2+}$, instead, the interplay of the $5d^1$–$4f$ emission transition and $^7F_{0,1} \rightarrow D_{0,1}$ absorption lines can change the emission pattern considerably depending on the temperature, crystal field strength, and splitting.[141] The configurational coordinate diagram of $Sm^{2+}$ in BaBrI is depicted in **Figure 3**, showing the thermal repopulation between $4f$ and $5d$ levels of $Sm^{2+}$, which can result in either the fast $5d \rightarrow 4f$ or slow $4f \rightarrow 4f$ emissions. The reason for using the $Sm^{2+}$ dopant consists of lower energy of its $5d$–$4f$ transitions, situated in the red/near-infrared spectral region, in the same host when compared with $Eu^{2+}$ or $Yb^{2+}$,[137] which enables efficient use of semiconductor detectors. Most of the research was focused on the exploration of the compositional space formed by ternary halides of alkali and alkaline earth metals, that is, $A_x^+ B_y^{2+} X_{x+2y}^-$ (where A stands for an alkali metal or thallium, B for alkaline earth metal, and X for halogen). Alkaline earth metals are isovalent with $Eu^{2+}$ ($Yb^{2+}$, $Sm^{2+}$) and have similar ionic radii (1.00, 1,18, 1,35, and 1.17 Å for $Ca^{2+}$, $Sr^{2+}$, $Ba^{2+}$, and $Eu^{2+}$, respectively, in VI

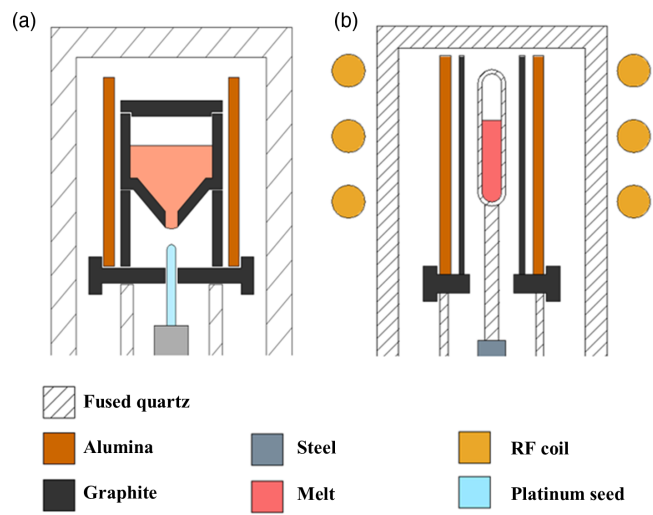

**Figure 2.** Scheme of halide crystal growth by a) μ-PD and b) mVB methods.

Legend:
- Fused quartz
- Alumina
- Graphite
- Steel
- Melt
- RF coil
- Platinum seed

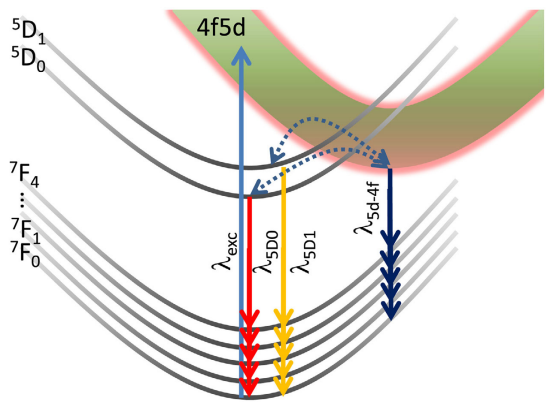

**Figure 3.** Configurational coordinate diagram of $Sm^{2+}$ in BaBrI. Solid arrows indicate radiative and dotted arrows non-radiative processes. Reproduced under terms of the CC-BY license.[141] Copyright 2020, The Authors, Published by Elsevier.

coordination[142]). This makes them the best candidates for $Eu^{2+}$ ($Yb^{2+}$, $Sm^{2+}$) substitution. Divalent lead cation is another suitable candidate for the $Eu^{2+}$ substitution. However, $Pb^{2+}$ is undesirable due to its toxicity. A suitable crystal lattice site is essential for the $Eu^{2+}$-activated scintillators, because of the high doping levels often required for optimal scintillation performance.[143] Several compositions with different A:B ratios form congruently melting compounds which can be explored for use as bulk single-crystal scintillators, namely, the $AB_2X_5$, $ABX_3$, $A_2BX_4$, $A_3BX_5$, and $A_4BX_6$ ones

The number of photons $N_{ph}$ emitted after absorption of a photon with energy $E_\gamma$ is given by the equation[144]

$$N_{ph} = \frac{E_\gamma}{E_g} YSQ \qquad (1)$$

where $E_g$ is the bandgap of the scintillator, parameters $Y$, $S$, and $Q$ are related to the conversion, transfer, and luminescence stages of scintillation mechanism, namely, $Y$ is the conversion efficiency, $S$ is the transfer efficiency, and $Q$ is the QL of the luminescent process. Due to the inverse correlation between the number of emitted photons and bandgap, materials with a narrow bandgap are prospective for achieving high light yield. Therefore, iodides were the most extensively investigated halide scintillators for high-LY applications.

### 3.1. $AB_2X_5$

All the compounds of interest from the $AB_2X_5$ family crystallize in $TlPb_2Cl_5$-type monoclinic ($P2_1/c$, space group no. 14) crystal structure with two crystallographic lattice sites of the $B^{2+}$ cation (coordination numbers 7 and 8). First-principle calculations predict a random distribution of $Eu^{2+}$ among the two $B^{2+}$ lattice sites in $ASr_2I_5$ compounds and a preferential occupation of the crystal site with CN 7 in $ABa_2I_5$ compounds.[145] The preferential occupation of $Eu^{2+}$ may result in the formation of Eu-rich domains in $ABa_2I_5$ crystals.[145]

$CsBa_2I_5$:$Eu^{2+}$ exhibits LY up to 102 000 ph MeV$^{-1}$[76] and energy resolution down to 2.3%[146] (at 662 keV $^{137}$Cs). Pulse height spectra are shown in **Figure 4**. Both the photopeak with ER 2.3% full width at half maximum (FWHM) and escape peak are well resolved. The $Sm^{2+}$ codoping of $CsBa_2I_5$:$Eu^{2+}$ was reported as the first attempt to prepare a "black scintillator" where an efficient $Eu^{2+}$–$Sm^{2+}$ energy transfer occurs, and in the energy of the resulting $5d^1$–$4f$ transition of $Sm^{2+}$, the emission peak is situated at around 755 nm. Scintillation response shows a high-energy resolution of 3.2% (662 keV $^{137}$Cs excitation) and a leading decay time of 2.1 μs.[147] The bromide analog $CsBa_2Br_5$:$Eu^{2+}$ also exhibits a very high LY of 92 000 ph MeV$^{-1}$ with an advantage of lower hygroscopicity.[148] $CsBa_2I_5$ is also a prospective material for activation with monovalent cations.[149] $KSr_2I_5$[150] and $KSr_2Br_5$[151] were introduced as other promising scintillators from the $AB_2X_5$ family showing high LY of 94 000 and 75 000 ph MeV$^{-1}$ and ER of 2.4% and 3.5% (at 662 keV $^{137}$Cs), respectively. These materials were further developed with regard to growth conditions,[71,152,153] $\gamma$ detection capabilities,[154,155] and mixed $KSr_2Br_xI_{5-x}$ compositions[156] resulting in high quality, large-volume (Ø1 × 1 in.) crystals for $\gamma$ spectrometry.[71] Results on rubidium[157] and barium[158] analogs were also reported to have similar parameters.

Substitution of an alkali metal with thallium results in better stopping power for high-energy $\gamma$ quanta due to higher $Z_{eff}$ while keeping high performance. Therefore, analogs based on chlorides,[159] bromides,[160] and iodides[70,161–163] were investigated. However, similarly to other $Eu^{2+}$-doped scintillators mentioned, the iodides exhibit the best performance with LY up to 72 000 ph MeV$^{-1}$[163] and energy resolution below 3% (at 662 keV $^{137}$Cs).[162,163]

### 3.2. $ABX_3$

Iodides from the $ABX_3$ family crystallize in either $NH_4CdCl_3$-type ($Pnma$, space group no. 62) or $UFeS_3$ ($Cmcm$, space group no. 63)-type orthorhombic crystal structure. Both crystal structures have a single octahedral $B^{2+}$ lattice site available for $Eu^{2+}$ doping. Scintillation properties of a large number of $Eu^{2+}$-doped crystals from the $ABX_3$ family were reported. However, most of the crystals does not exceed LY of 50 000 ph MeV$^{-1}$ and ER of 5% (at 662 keV $^{137}$Cs) except for $CsSrI_3$ and $KCaI_3$. For $CsSrI_3$, the LY value up to 65 000 ph MeV$^{-1}$[164] and ER down to 3.4%[165] (at 662 keV $^{137}$Cs) were reported. However, the growth of large-volume single crystals of CsSrI$_3$ has proven very complicated. KCaI$_3$ has emerged as the most promising candidate from the ABX$_3$ family having LY of 72 000 ph MeV$^{-1}$ and ER of 3%[55] for a small crystal (Ø 2 mm). Further improvement was achieved via Sr substitution.[166,167] Moreover, codoping with RE$^{3+}$ results in significant suppression of the afterglow,[168] while Zr$^{4+}$ codoping results in improvement of energy resolution.[169] Growth of large volume crystals (up to Ø50 mm) is possible without significant deterioration of the scintillation performance.[61] The Yb$^{2+}$-doped ACaCl$_3$ (A = Cs, Rb, K) crystals show emission bands centered at ≈445 nm with two decay time constants of ≈10 and 200 μs in the PL and scintillation decays attributed to fast (LS) and slow (HS) 5d–4f transition of Yb$^{2+}$, respectively.[170] LY values of ACaCl$_3$:Yb$^{2+}$ were estimated at about 3,600–5,200 ph MeV$^{-1}$ based on the pulse height spectra, that is., much lower compared with the aforementioned KMeI$_3$:Eu (Me=Ca, Sr), which can be explained by the presence of a very slow decay component due to the involvement of the HS 5d state of Yb$^{2+}$ in the decay mechanism.

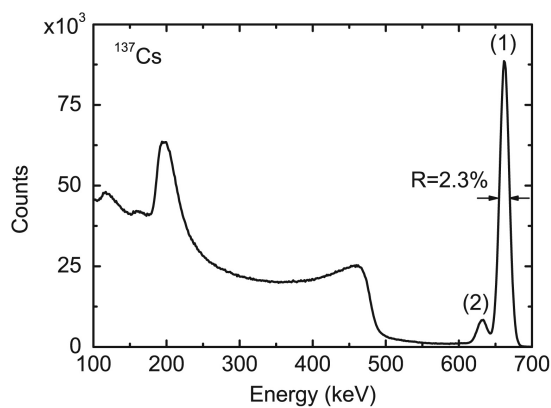

**Figure 4.** Pulse height spectrum of $CsBa_2I_5$:Eu(0.5%). Reproduced with permission.[146] Copyright 2013, Elsevier.

### 3.3. A$_4$BX$_6$

All the compounds of interest from the A$_4$BX$_6$ family crystallize in the K$_4$CdCl$_6$ type (R-3c, space group no. 167) with a single octahedral B$^{2+}$ lattice site. The scintillation properties were reported for Cs$_4$CaI$_6$ and Cs$_4$SrI$_6$ with LY up to 69 000 and 71 000 ph MeV$^{-1}$ and ER down to 3.6% and 3.2% (at 662 keV $^{137}$Cs) respectively.[171,172] The effect of alkali metal substitution[172] was investigated. In the case of Yb$^{2+}$ activator, the best scintillation performance was achieved with Cs$_4$CaI$_6$:Yb (1%$_{mol}$), which had a 3.5% energy resolution at 662 keV and 43 000 ph MeV$^{-1}$ light yield.[173] Lower light yield compared wuth the Eu$^{2+}$-doped Cs$_4$CaI$_6$ might be due to worse timing characteristics of the scintillation response because of the involvement of HS 5d level of Yb$^{2+}$ as mentioned above. Moreover, this matrix allows complete substitution of B$^{2+}$ cation with Eu$^{2+}$. Single crystals of Cs$_4$EuBr$_6$ and Cs$_4$EuI$_6$ exhibit LY of 78 000 and 53 000 ph MeV$^{-1}$ and ER of 4.3% and 5.0% (at 662 keV $^{137}$Cs) respectively.[174]

### 3.4. SrI$_2$

Even though SrI$_2$:Eu was patented by Hofstadter,[51] it regained the interest of the scientific community at the beginning of the last decade due to technological advancements that allowed large-volume high-quality single-crystal growth of this high-performance scintillator. In the last decade, the R&D of SrI$_2$ was focused on the improvement of its properties via codoping,[66,175,176] optimizing the growth conditions,[65,103,177,178] and packaging.[179] Moreover, Yb$^{2+}$ doping was also suggested as prospective due to the faster scintillation kinetics.[180]

## 4. Ce$^{3+}$-Activated Fast Scintillators

Similarly to the Eu$^{2+}$-activated halide scintillators, the search for new Ce$^{3+}$-activated halide ones was focused on the exploration of compositional space formed by ternary halides involving alkali metals (or thallium). The motivation for Ce$^{3+}$ doping is due to its at least 20 times faster luminescence lifetime compared with Eu$^{2+}$ even if it is based on the same type of 5d-4f transition and somewhat bigger Stokes shift which decreases unwanted reabsorption. In the case of Ce$^{3+}$ activation, the alkali earth metals were replaced by RE metals to create a crystal lattice site suitable for Ce$^{3+}$ doping.

It is worth explaining here an absence of reports in bromide and iodide hosts dealing with Pr$^{3+}$ emission center and its 5d–4f luminescence which is typically two times faster compared with that of Ce$^{3+}$. The reason is that in bromides and iodides the Pr$^{3+}$ ground state becomes immersed in the valence band, see **Figure 5**, which results in its preferential excitation through the charge transfer absorption band followed with emission from the lower-lying 4f $^3$P$_1$ state, which occurs with the lifetime of the order of ten microseconds. Even in the chloride host, namely, LuCl$_3$, where both the 4f ground-state $^3$H$_4$ and excited-state 5d$^1$ are located in the forbidden gap and the transition 4f($^3$H$_4$)–5d$^1$ of Pr$^{3+}$ is located at lower energy compared with the charge transfer one, the band-to-band excitation results in energy transfer from the exciton state right to lower-lying 4f $^3$P$_1$ and $^1$D$_2$ levels and their much slower luminescence so that the fast 5d$^1$–4f radiative transition is avoided.[181]

### 4.1. CeBr$_3$

To this date, lanthanum bromide doped with cerium and codoped with strontium (LaBr$_3$:Ce,Sr) exhibits the best energy resolution (2.0% at 662 keV of $^{137}$Cs) among bulk single-crystal inorganic scintillators.[182] However, the attention in R&D shifted from LaBr$_3$:Ce to CeBr$_3$ due to internal contamination with $^{138}$La (0.089% abundance, $T_{1/2} = 1.05 \cdots 10^{11}$ y) radionuclide and a high tendency for cracking during crystal growth. Based on the knowledge of LaBr$_3$:Ce, the influence of aliovalent doping (mainly by alkaline earth metals) was investigated,[183–185] suggesting that codoping with Sr$^{2+}$ results in the best energy resolution in both LaBr$_3$:Ce and CeBr$_3$.[184] The influence of codoping on the nonproportionality of the scintillation response in CeBr$_3$ is illustrated in **Figure 6**. Moreover, the influence of CeBr$_{3-x}$Cl$_x$[186] or CeBr$_{3-x}$I$_x$[187] solid solutions on scintillation properties was investigated.

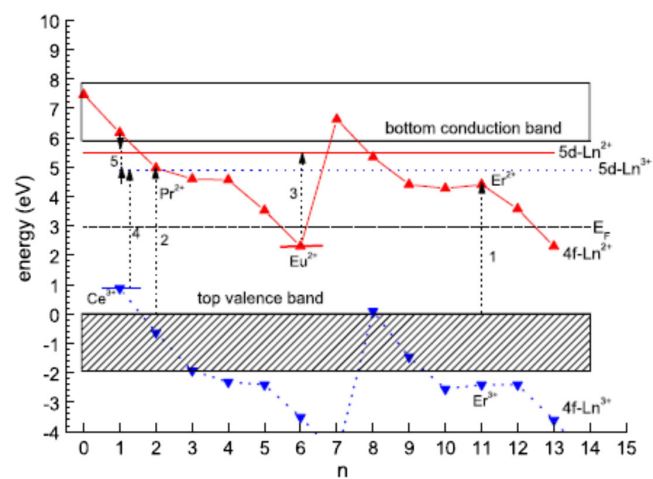

**Figure 5.** Location of divalent and trivalent lanthanide energy levels in LaBr$_3$ as a function of the number n of electrons in the 4f shell of the trivalent lanthanide. Note, n is the number of electrons in the 4f shell of the divalent lanthanide. Reproduced with permission.[354] Copyright 2005, Elsevier.

### 4.2. A$_2$A'LnX$_6$

The successful development of Cs$_2$LiYCl$_6$:Ce (CLYC) as a thermal neutron detector with excellent $n^0/\gamma$ discrimination[65] initiated extensive research in scintillators from the elpasolite family. These materials offer a combination of high proportionality, resulting in a good energy resolution and a suitable site for Ce$^{3+}$ doping. Moreover, cesium-containing chloride elpasolites exhibit crossluminescence which can be used for discrimination of neutron/alpha radiation from beta and gamma one.

Substitution of Cs with Tl results in a material with higher stopping power while keeping comparable energy resolution and light yield.[188,189] However, the substitution of chlorine with heavier halides results in improvement of the light yield and energy resolution.[190–193] However, heavy halide analogs of CLYC do not exhibit fast CL emission. First-principle calculations suggest that Ag substitution on A' position should further improve scintillation properties.[194]

### 4.3. A$_3$LnX$_6$

Materials from this family are mostly monoclinic with either K$_3$MoCl$_6$ (P2$_1$/c, space group no. 14) or Cs$_3$BiCl$_6$ (C2/c, space group no. 15)-type crystal structure at RT. Both have a single octahedral Ln$^{3+}$ lattice site. This often results in complicated crystal growth due to phase transitions connected with the lowering of crystal symmetry during cooling.[127,195] Most of the investigated compositions contained La or Gd on Ln position and Cs at A position.[192,195,196] The best results were reported for Cs$_3$GdBr$_6$:Ce with LY of 47 000 ph MeV$^{-1}$ and ER of 4.0% (at 662 keV $^{137}$Cs).[195] Full substitution with Ce at Ln position is possible, however, with inferior performance.[127,197] Lastly, Li$_3$YCl$_6$:Ce was suggested as a thermal neutron scintillator.[198]

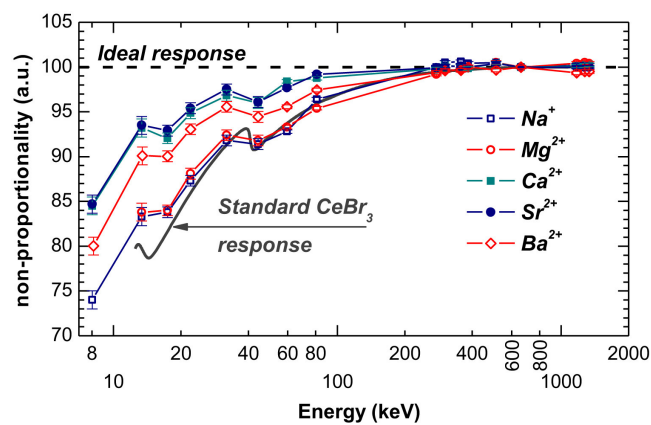

**Figure 6.** Measured nonproportionality characteristics of standard and codoped CeBr$_3$ samples. For all samples, the values are normalized to 100% at 662 keV. Reproduced with permission.[184] Copyright 2013, Elsevier.

### 4.4. A$_2$LnX$_5$

All compounds of interest from the A$_2$LnX$_5$ family crystallize in the orthorhombic crystal structure (Pnma, space group no. 62) with K$_2$PrCl$_5$-, Cs$_2$DyCl$_5$-, or Y$_2$HfS$_5$-type structure with a single Ln$^{3+}$ lattice site. Compositions with Tl on the A site exhibit the density up to 5.98 g cm$^{-3}$ and $Z_{eff}$ up to 79.[199] Therefore, Tl$_2$LaCl$_5$,[9,70,199–202] Tl$_2$LaBr$_5$,[199,202,203] Tl$_2$GdCl$_5$,[204,205] and Tl$_2$GdBr$_5$[70] were most extensively investigated. Best samples reported in the literature exhibit LY up to 82 000 ph MeV$^{-1}$[201] and ER down to 2.8% (at 662 keV $^{137}$Cs).[202] Crystals with complete Ce$^{3+}$ substitution on the Ln$^{3+}$ site exhibit inferior performance similar to A$_3$LnX$_6$ compounds.[206]

### 4.5. ALn$_2$X$_7$

Most of the compositions from the ALn$_2$X$_7$ family crystallize in either KDy$_2$Cl$_7$-type monoclinic (P2$_1$/c, space group no. 14) or RbDy$_2$Cl$_7$-type orthorhombic (Pnma, space group no. 62) crystal structure. Both have two Ln$^{3+}$ lattice sites. However, recently the crystal structure of CsCe$_2$Cl$_7$ was derived from single-crystal diffraction having a new, previously unreported, monoclinic crystal structure.[207] Most of the reported crystals from the ALn$_2$X$_7$ family have low quality due to complicated crystal growth[126,208–210] which results in worse scintillation performance. However, gadolinium-containing crystals were suggested as scintillators for both gamma and thermal neutron detection.[211]

## 5. Activation with ns$^2$ Ions

The first member of the so-called heavy 6s$^2$ ion group, the Tl$^+$ ion, was used as the luminescence center in the very first single-crystal halide scintillator NaI:Tl in the late 1940s.[212] Two closely spaced $^3$P$_0$ and $^3$P$_1$ levels constitute its lowest excited state. Given the strong spin–orbit coupling, the $^3$P$_1$ shows the decay times of the order of a few hundreds of nanoseconds which is faster compared with the 5d–4f transition of Eu$^{2+}$. Furthermore, a considerably higher Stokes shift completely removes the reabsorption risk even in large scintillation elements. The Tl$^+$ has been recently adopted even as a cation in ternary iodides as mentioned above and the electronic band structures of InBa$_2$I$_5$ or TlBa$_2$I$_5$ were calculated showing noticeable contribution from In$^+$ and Tl$^+$ in the valence and conduction band edges.[213] Tl$^+$ and In$^+$ were also doped in CsBa$_2$I$_5$[149] providing LY values of 40 000 and 35 000 ph MeV$^{-1}$, respectively. Lower toxicity of indium is an advantage, but its weaker spin–orbit coupling results in 5–10 times slower luminescence and scintillation response compared with Tl$^+$ so that the dominant component in the decay is of the order of a few microseconds.

## 6. Self-Activated/Intrinsic Scintillators

### 6.1. A$_2$MX$_6$

Many compositions with formula A$_2$MX$_6$ where A stands for alkali metal (or thallium) and M stands for tetravalent cation

(mostly Hf and Zr) crystallize in the cubic $K_2PtCl_6$-type structure (*Fm-3m*, space group no. 225). The crystal structure often called "vacancy-ordered double perovskite" can be derived by doubling the $ABX_3$ perovskite structure with one $B^{2+}$ cation replaced by $M^{4+}$ cation and the other by a vacancy, resulting in charge neutrality of the structure. Compositions with smaller alkali metal (i. e. Rb, K, or Na) on the A lattice site diverge from the cubic crystal structure and undergo distortion to the lower tetragonal or monoclinic symmetry.[214] Even though the luminescence of $Cs_2HfX_6$ was investigated by Ackerman already in 1984,[215] this material family did not attain much attention of the scintillation community until its rediscovery by Burger et al. in 2015.[8]

$Cs_2HfCl_6$ is prospective due to the combination of high $Z_{eff}$, high LY of 54 000 ph MeV$^{-1}$,[8] and high proportionality resulting in excellent energy resolution of 3.3% (at 662 keV $^{137}Cs$)[8] in an undoped crystal with low hygroscopicity. The origin of intrinsic luminescence in $Cs_2HfCl_6$ was ascribed to radiative recombination of excitons trapped on $[HfCl_6]^{2-}$ octahedra.[72,216] The emission mechanism is schematically sketched in **Figure 7**.

These promising results motivated investigation of the properties of $Cs_2HfCl_6$ using various methods including simultaneous thermogravimetric analysis,[217] electron paramagnetic resonance (EPR), and thermally stimulated luminescence (TSL).[72,73,218] Moreover, improvement of material purification and crystal growth conditions[121,219,220] resulted in improvement of the ER down to 2.8% (at 662 keV $^{137}Cs$).[219] Furthermore, analogs of $Cs_2HfCl_6$ were investigated including $A_2MCl_6$,[221,222] $Cs_2HfBr_xCl_{1-x}$,[45,223–225] $A_2HfI_6$,[226–228] $Tl_2MCl_6$,[199,229–232] and $Tl_2HfBr_6$.[199] All exhibit broadband emission with scintillation decay times in the microsecond range. The progress in the development of several scintillators with $K_2PtCl_6$ structure was recently reported by Hawrami et al.[233] Substitution of chlorine with heavier halogens results in faster decay times due to stronger spin–orbit coupling[216] and higher LY due to a decrease in the bandgap. Moreover, a redshift of the self-trapped exciton (STE) emission can be utilized to match the spectral sensitivity of the avalanche photodiode[120,226] or region of low-transmission losses of silica optical fibers.[228] However, bromide and iodide analogs of $Cs_2HfCl_6$ suffer from high hygroscopicity. For a review of the $Cs_2HfCl_6$-type scintillators, see the study by Nagorny.[234] First attempts of preparation of ceramic $Cs_2HfCl_6$[235,236] and $Tl_2HfCl_6$[236] showed very promising results.

Another advantage of scintillators from the $A_2MX_6$ family is excellent pulse-shape discrimination of scintillation response to α and γ radiation.[237,238] This makes $Cs_2HfCl_6$ (and other scintillators from this family) prospective for the detection of fast neutrons in mixed neutron–gamma radiation fields. Moreover, the combination of α/γ discrimination together with high Hf concentration allowed measurement of the rare α decay time of $^{174}Hf$ in the "source=detector" configuration.[239] However, such applications require the use of highly radiopure starting chemicals.[237]

Activation of the $A_2MX_6$ matrix with $Tl^+$,[240] $Eu^{2+}$,[241] and $Ce^{3+}$[240,241] was unsuccessful so far, as well as doping with alkaline Earth metals.[242] However, doping on the $Hf^{4+}$ site with either $Zr^{4+}$[243] or $Te^{4+}$[244] results in a redshift of the emission.

### 6.2. $Cs_3Cu_2I_5$ and $CsCu_2I_3$

Ternary cesium copper iodides have gained significant attention from the optoelectronic community due to their nontoxicity, air stability, and PLQY up to 91.2% for $Cs_3Cu_2I_5$ single crystals.[10] $Cs_3Cu_2I_5$ crystallizes in orthorhombic "0D perovskite" crystal structure (*Pnma*, space group no. 62) consisting of separated $[Cu_2I_5]^{3-}$ units separated by $Cs^+$ cations.[53] The first report of the bulk single-crystal $Cs_3Cu_2I_5$ showed promising results with LY of 18 000 ph MeV$^{-1}$ and ER of 7.7% (at 662 keV $^{137}Cs$) for undoped crystal[74] and a significant improvement of scintillation performance after Tl doping. The sample doped with 1% of Tl exhibited LY of 51 000 ph MeV$^{-1}$ and ER of 4.5% with an excellent proportionality.[74] Improvement in crystal quality increased LY up to 41 500 ph MeV$^{-1}$[53] and ER down to 3.6%[245] for an undoped crystal. Furthermore, Tl-doped crystals exhibit LY close to 100 000 ph MeV$^{-1}$[53] and ER below 3.5%.[53,246] This matrix also shows an extremely low afterglow of 0.015% (10 ms after excitation cut-off) for 0.1% Tl-doped samples.[246] Based on scintillation kinetics[74] and first-principle calculations,[53] the scintillation mechanism in $Cs_3Cu_2I_5$ was not attributed to ns→s$^2$ radiative transition of $Tl^+$, but the recombination of a self-trapped hole with an electron trapped at the Tl impurity. The scintillation mechanism in Tl-doped $Cs_3Cu_2I_5$ is schematically illustrated in **Figure 8**.

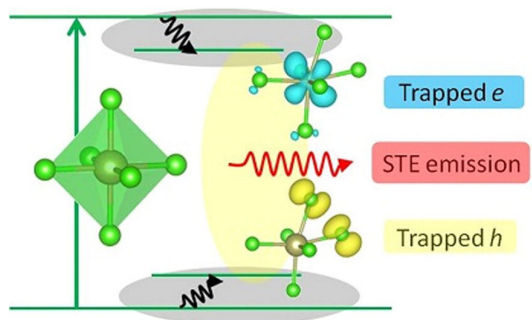

**Figure 7.** Schematics of luminescence mechanism in $Cs_2HfCl_6$. Reproduced with permission.[216] Copyright 2016, American Chemical Society.

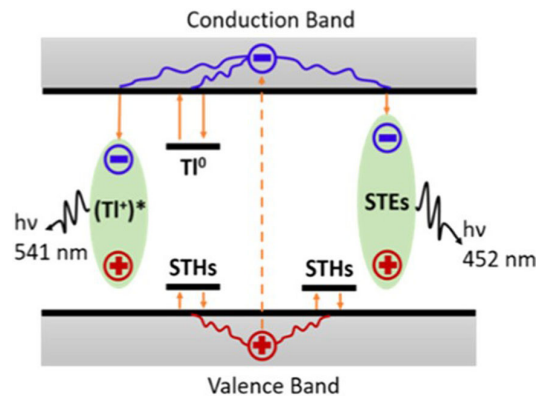

**Figure 8.** Schematic illustration of scintillation mechanism in Tl-doped $Cs_3Cu_2I_5$. Reproduced with permission.[74] Copyright 2020, American Chemical Society.

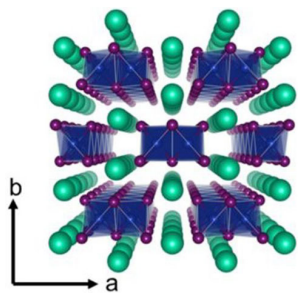

**Figure 9.** View of the CsCu$_2$I$_3$ crystal structure along the *c*-axis. Reproduced with permission.[247] Copyright 2021, American Chemical Society.

CsCu$_2$I$_3$ crystallizes in orthorhombic "1D perovskite" crystal structure (*Cmcm*, space group no. 63) consisting of chains of [Cu$_2$I$_6$]$^{4-}$ edge-sharing octahedra separated by rows of Cs$^+$ cations.[247] **Figure 9** depicts a view of the crystal structure along the *c*-axis. Under X-ray excitation, CsCu$_2$I$_3$ exhibits broadband emission peaking at 570 nm which was ascribed to localized exciton emission.[247] LY and ER at RT are inferior to that of Cs$_3$Cu$_2$I$_5$. However, CsCu$_2$I$_3$ exhibits an extremely low afterglow at the level of 0.008% (10 ms after excitation cut-off).[247] Moreover, the temperature dependence of PL revealed the potential of CsCu$_2$I$_3$ to achieve LY above 100 000 ph MeV$^{-1}$ via exciton engineering.[247]

## 7. Cross-Luminescence Scintillators

Cross-luminescence (CL, also called "Auger-free luminescence" or "core-valence luminescence") is radiative recombination of an electron from the valence band with a hole from the uppermost core band. It is observed mostly in halides containing heavier alkali metals (i.e., K, Rb, and Cs) or barium (see the review in another study[248]). A band structure of BaF$_2$ with an indication of both the CL and STE emissions is schematically shown in **Figure 10**. CL is prospective for fast timing applications due to fast decay kinetics with decay time ≈1 ns. However, the position of CL emission in vacuum ultraviolet (VUV) complicates the practical application of CL scintillators. The recent development of UV-sensitive semiconductor photodetectors has renewed interest in CL scintillators. Using VUV-HD silicon photomultiplier (SiPM), a coincidence time resolution (CTR) of 51 ps was achieved for commercially available BaF$_2$ crystal.[249] Currently, this is the lowest CTR value reported for the inorganic scintillator. Cross-luminescence can also be exploited for $\alpha(n^0)$ and $\gamma$ discrimination.[250]

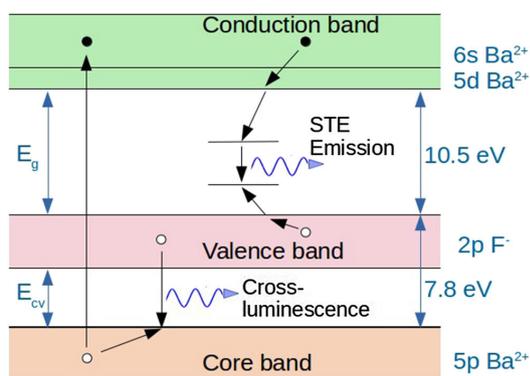

**Figure 10.** The band structure schematics of BaF$_2$ showing CL and STE emission. Reproduced under terms of the CC-BY license.[355] Copyright 2020, The Authors. Published by Frontiers.

An ideal cross-luminescence scintillator should exhibit the emission redshifted compared with BaF$_2$ while keeping the fast scintillation kinetics and relatively high LY (≈1400 ph MeV$^{-1}$ for BaF$_2$[251]) without major slow components. The redshift of CL emission can be achieved by exploring cesium-based fluorides and chlorides. In the last decade, the research was focused on chloride materials, such as CsBCl$_3$ (B=Mg, Ca, Sr),[252–255] Cs$_2$BaCl$_4$,[255,256] Cs$_2$ZnCl$_4$,[257] and Cs$_3$ZnCl$_5$.[257] Preliminary measurements of time resolution of CsCaCl$_3$ reported CTR of 148 ps[255] with the potential to surpass BaF$_2$. Only a few reports on CL in single-crystal fluorides can be found in the recent literature.[258,259]

## 8. Transparent Ceramics

Transparent ceramic scintillators were first introduced in the 1980s.[260,261] In the following years, the research was focused on oxide materials, mainly of the garnet structure. For more information about oxide ceramic scintillators, see the study by Nikl et al.[262] and references therein. The first attempt for preparation of the halide transparent ceramic scintillator was reported by Wisniewski[263] in 2007. They used hot pressing to produce uniformly shaped, millimeter-size translucent LaBr$_3$:Ce ceramics with an estimated LY of 42 000 ph MeV$^{-1}$.[263,264] In the last decade, development of translucent/transparent ceramics was reported for several halide scintillators, including BaCl$_2$,[265] Ba$_{1-x}$La$_x$Cl$_{2+x}$,[266] SrI$_2$:Eu,[267] Cs$_2$HfCl$_6$,[236,263] and Tl$_2$HfCl$_6$.[236]

## 9. Microstructured Halide Scintillators

Microstructured scintillators were developed for imaging screens to enable an increased X-ray stopping power with the help of increased thickness of scintillation element and simultaneously keeping high spatial resolution enabled by its light-guiding property. The columnary grown CsI:Tl became commercially successful[268–270] prepared by vacuum evaporation in the form of long (up to 1 mm) and thin (from few to several μm in diameter) densely packed needles, which are reasonably optically isolated. In practical applications, such a needle layer can be deposited directly on a semiconductor photodetector.[3] Radiographic panels of this kind were developed in the late 1980s and became widely used in routine radiographic imaging.[271]

Another approach in the scintillator element structure design appeared about a decade ago, namely, the so-called phase-separated scintillators, prepared from an eutectic melt by unidirectional solidification (using most frequently the Bridgman, CZ, or micro-pulling-down techniques).[272] Such a manufacturing technique enables the preparation of a monolithic element consisting of a crystal host with the embedded ordered thin rod-like second phase, as shown in **Figure 11**.

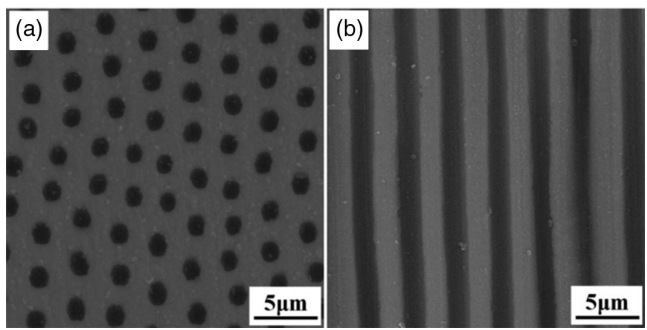

**Figure 11.** SEM images of the CsI–NaCl system with light and dark visual appearance, respectively. a) Top view and b) cross-sectional view. Reproduced with permission.[272] Copyright 2012, John Wiley and Sons.

Light-guiding property is achieved in the constituent of higher refraction index which usually works as a scintillator. In principle, both the host crystal and rod-like second phase can work in such a way, though light guidance within tiny rods is more efficient. In the study by Yasui et al.,[272] a couple of alkali halide compounds were used to demonstrate such a light-guiding property in scintillation mode and as the most advantageous eutectic composition the CsI(host)-NaI:Tl(scintillator) was found. It is worth noting that in this case, the higher refraction index is that of CsI, so that the generated scintillation light is guided out through the host constituent and not through the rod-like NaI:Tl scintillator phase. Scintillation characteristics of eutectic composition from this material family were reported later also in other studies,[87,273,274] focusing on a possibility to prepare larger optical elements for imaging devices; see an example of the grown Tl-doped CsI/CsCl/NaCl eutectic in **Figure 12**. In this case, the LY of the 12 mm h$^{-1}$-pulled sample was about 16 000 ph MeV$^{-1}$, that is, 27% of the CsI:Tl standard single-crystal scintillator. The scintillation decay time value was measured to be 910 ns, the same as that of the CsI:Tl standard scintillator. The performed imaging test with β particles excitation and charged-coupled device (CCD) photodetectors provided a position resolution of 16 μm FWHM.

Frequently, fluoride-based eutectic scintillators were reported based on the Eu$^{2+}$-doped $^6$LiF/CaF$_2$,[275] or $^6$LiF–SrF$_2$,[276] Ce-doped LiF–SrF$_2$,[277] or undoped LiF–SrF$_2$ and LiF–CaF$_2$[278] where the emission is due to the exciton in the alkali earth fluoride constituent. Eutectic composition of LiF host and scintillator phase based on another doped fluoride was reported for Eu$^{2+}$-doped LiF/CaF$_2$/LiBaF$_3$[279] and LiF/LiBaF$_3$,[280] and Ce$^{3+}$-doped fluorides LiF/LaF$_3$,[280] LiF/LiLuF$_4$,[281] and LiF/LiYF$_4$.[282] From the point of view of efficient light guidance, the LiF/LaF$_3$ composition appears the most promising due to the largest difference of refractive indexes ($n_{LiF} = 1.39$, $n_{LaF3} = 1.60$ at the emission peak of Ce$^{3+}$), which enhances the total reflection in the rod-like LaF$_3$:Ce scintillator. Given the LiF host, such materials were considered also for thermal neutron detection, especially in the case of $^6$Li enrichment which provides a large energy deposit due to the reaction

$$n + {}^6\text{Li} \rightarrow {}^3\text{H}(2.75\,\text{MeV}) + {}^4\text{He}(2.05\,\text{MeV}) \qquad (2)$$

As the smaller bandgap of scintillation phase may provide higher efficiency due to smaller energy required to create an electron–hole pair, apart from the reports dealing with chloride eutectic compositions, such as LiCl–BaCl$_2$:Eu$^{2+}$,[283] LiCl/Li$_2$SrCl$_4$:Eu$^{2+}$, and[69] LiCl–CeCl$_3$,[284] those based on bromides, such as Ce:LaBr$_3$/AEBr$_2$ (AE = Mg, Ca, Sr, Ba)[68] and LiBr/LaBr$_3$:Ce,[285] or iodides as BaI$_2$/LuI$_3$:Ce[286] or Eu-doped LiSrI$_3$/LiI,[287] were published as well. The highest LY was reported in,[285] 74 000 ph per neutron, which is much superior to presently available commercial single-crystal scintillators for thermal neutron detection. A practical drawback of most of these materials is, however, their hygroscopicity so that the hermetic sealing of the eutectic element becomes necessary.

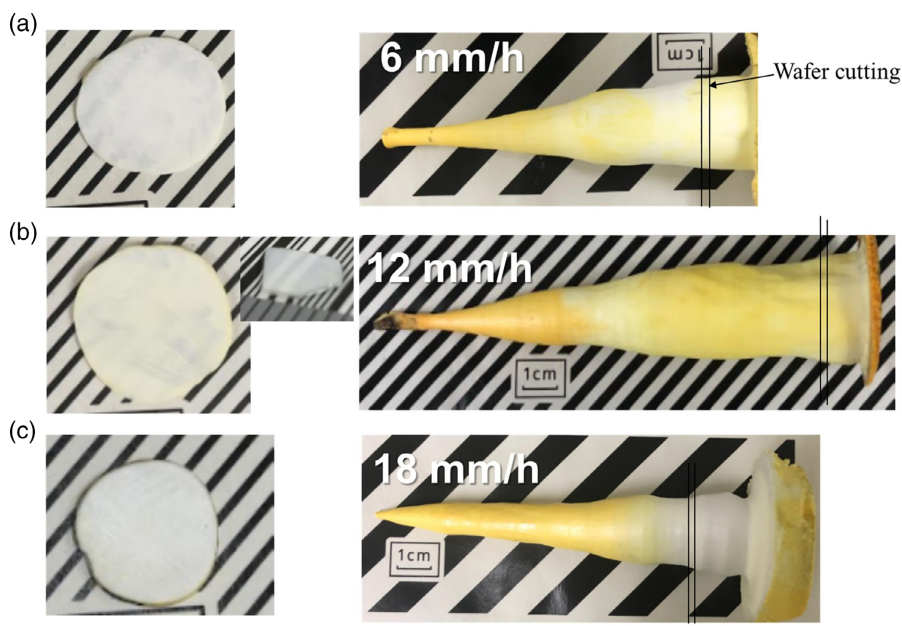

**Figure 12.** Photographs of the grown Tl-doped CsI/CsCl/NaCl eutectic (right) and the wafers cut in the transverse cross section (left) for the pulling rate of a) 6, b) 12, and c) 18 mm h$^{-1}$. Reproduced with permission.[87] Copyright 2021, Elsevier.

## 10. Halide Perovskite NCs

### 10.1. Hybrid Perovskites

Increased interest in research of halide perovskites first focused on the family of organic–inorganic, or so-called hybrid, perovskites, where the cuboctahedrally coordinated A site of the $ABX_3$ perovskite formula is occupied by organic cations, such as $CH_3NH_3$ (methylammonium: MA) and $NH_2CH=NH^{2+}$ (formamidinium: FA). B cations, octahedrally coordinated by X anions, are metal atoms, typically $Pb^{2+}$ or $Sn^{2+}$, and X anions are usually Cl, Br, or I. Hybrid perovskites became drivers of alternative photovoltaic technology in the past decades. In the early work,[288] $MAPbBr_3$ and $MAPbI_3$ were used as light absorbers in photovoltaic devices. Since then, the bulk and nanocrystalline hybrid perovskites were mainly investigated in the solar cell applications (see, e.g., a recent review[289]). The certified efficiency of perovskite solar cells has increased drastically to over 25% and can now compete in this respect with Si technology.[290]

Recently, hybrid perovskites in nanocrystalline form were investigated as potentially interesting candidates also for scintillator applications. High LY alpha particle scintillators based on hybrid perovskite $MAPbBr_3$ NCs (NCs) with an ultrafast response time at low temperatures are reported.[291] The LY of the hybrid perovskite $MAPbBr_3$ increases with decreasing temperature and appears much higher when compared with that of the commercial $(Lu,Y)_2SiO_5$:Ce (LYSO:Ce)-based scintillators at around 150 K as well as CsI-based scintillators at around 50 K. In addition, it also showed an ultrafast alpha response with the fastest decay component of 0.1 ns at 77 K. The measurement of the pulse height spectra showed that $MAPbBr_3$ exhibits an ability to detect alpha particles comparable with that of the commercial scintillators.

$FAPbBr_3$ colloidal NCs were investigated for fast neutron imaging through the detection of recoil protons generated by neutron scattering.[292] A variety of NCs were screened for their LY and spatial resolution under fast neutron irradiation, with $FAPbBr_3$ providing the highest LY. Concentration and thickness-dependent measurements reveal that self-absorption and low concentrations are the primary limiting factors in these scintillators, providing design principles to foster the development of next-generation fast neutron detectors based on colloidal semiconductor NCs.

### 10.2. All-Inorganic Halide Perovskites

In all-inorganic halide perovskites, the A site of the $ABX_3$ formula is occupied by a monovalent $Cs^+$ cation. Colloidal NCs represent highly promising types of semiconductors for a variety of applications due to their narrow band photoluminescence (PL), tunable from ultraviolet to infrared spectral range, by simply manipulating the NC size and halide composition. High quantum efficiency and (ultra) fast decay times became attractive features also for manufacturing a new generation of scintillation detector systems. A body of associated work can be found in recent reviews, those mentioned in the Introduction or also other studies.[293,294] In the following we focus on some of the most recent works.

The most critical issue for scintillator performance can be carrier trapping. The study[295] shows $CsPbBr_3$ NCs as superior candidates for scintillating materials compared with higher-dimensional analogs. $CsPbBr_3$ with increasing dimensionality, namely, nanocubes, nanowires, nanosheets, and bulk crystals, was comprehensively studied to shed light on trapping and detrapping mechanisms to and from shallow and deep traps. The study involved radioluminescence and PL measurements together with thermally stimulated luminescence and afterglow experiments. All systems show detrapping physics related to shallow localized states due to surface defects. However, unlike in NCs, in higher-dimensional analogs, the suppression of defects needs to be operated on both the surface and volume levels.

Target applications of inorganic halide perovskites in the scintillation detection systems include X-ray imaging (see some works mentioned in Subsection 10.4. Nanocomposites), fast neutron imaging, and fast timing applications.

Successful fast neutron imaging scintillator requires a compelling and rare combination of high PLQY, high concentrations of NCs, and high Stokes shifts with low self-absorption. Such a combination was only recently found in colloidal NC system of $Mn^{2+}:CsPb(Br,Cl)_3$. A new synthesis based on zwitterionic ASC18 ligands yielded NCs which are colloidally stable to high concentrations while maintaining excellent optical quality. Fast neutron imaging experiments showed that these dense ASC18-capped NCs achieved light yields over 8 times greater than those of their oleyl-capped counterparts. Furthermore, thickness and concentration-dependent measurements under fast neutron irradiation showed the lack of self-absorption in this doped system, with essentially linear concentration dependence and drastically enhanced scalability with greater than 91.9% of the expected LY achieved for a tenfold increase in thickness.[296]

Fast timing applications, in particular those in medical imaging (time-of-flight-positron emission tomography [TOF–PET]) and high-energy physics, put challenging requirements especially on the timing resolution of scintillating material. A possible way to overcome the limits of standard scintillators so far commonly used is based on the metascintillator concept,[297] combining and optimizing several functionalities in the same scintillator heterostructure. In such heterostructures, the semiconductor NCs with ultrafast decay times, such as lead halide perovskites, can serve as efficient time taggers.

Three different perovskite NCs, $CsPbBr_3$, $FAPbBr_3$, and $CsPbI_3$, were investigated in the study by Maddalena.[298] Temperature-dependent RL measurements revealed that all NCs exhibit intense emission at cryogenic temperatures. $CsPbBr_3$ NCs exhibit negative temperature quenching (increasing emission with increasing temperature), leading to a high RL emission even at RT, with a deterministic LY of $24\,000 \pm 2100$ ph MeV$^{-1}$. On the other hand, $FAPbBr_3$ and $CsPbI_3$ NCs exhibit thermal quenching, leading to a lower RL emission at RT. The materials also exhibit very small afterglow with no traps. Investigated NCs display a fast RL decay time, with the average decay times below 6 ns for $CsPbBr_3$ and $CsPbI_3$ NCs and below 20 ns for FAPbBr$_3$ NCs, faster than those of previous CdSe/ZnS NCs,[299] cadmium telluride (CdTe) NCs,[300] and commercial scintillators that are currently used for fast-timing applications.

In the study by Kang et al.,[301] the authors study the time-resolved RL response of CsPbBr$_3$ QDs in their clustered state induced by $\alpha$ particles. Based on the statistical analysis of timing characteristics, they assessed the potential of using perovskite nanomaterials in timing applications performing a comparative measurement with a CsI(Tl) scintillator. The RL yield and detection efficiency were estimated to be 2.95 ph keV$^{-1}$ and 29.2%, respectively, referring to the mean cluster thickness of 5 QD layers.

Lead halide perovskite NCs with 3D structure have direct bandgap[302] featuring one main PL peak.[18] Due to toxicity of Pb element and poor stability of lead halide perovskite NCs in air, a great deal of effort was devoted to develop lead-free alternatives with improved stability. Lead-free variants based on Bi$^{3+}$, Sb$^{3+}$, or Sn$^{2+}$ and lead-free double perovskites, such as Cs$_2$AgBiBr$_6$ or Cs$_2$AgInCl$_6$, have attracted research attention. Interestingly, these reported lead-free perovskite NCs generally show an indirect-bandgap character. Consequently, the coexistence of direct and indirect transitions in some lead-free perovskite NCs could potentially induce the dual-color emission. For more details on lead-free perovskite NCs see, e.g., the recent review.[303]

### 10.3. Synthesis and Upscaling

The first synthesis that was introduced specifically for the colloidal CsPbX$_3$ NCs was hot injection.[18] This method is based on the precipitation of degassed precursors in an inert atmosphere at high temperatures (150–200 °C). This allows for good control of the reaction conditions and therefore good reproducibility. This method has been readily adapted for other shapes of NCs.[304–307]

First, cesium precursor (cesium oleate, Cs-ol) is synthesized. Degassed oleic acid is mixed with cesium carbonate in degassed 1-octadecene and the mixture is heated to 150 °C. Cs-ol is formed according to reaction (3) (-ol refers to the deprotonated form of oleic acid CH$_3$(CH$_2$)$_7$CH=CH(CH$_2$)$_7$COO$^-$).[308]

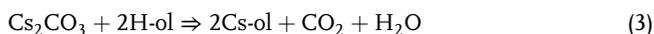

$$Cs_2CO_3 + 2H\text{-ol} \Rightarrow 2Cs\text{-ol} + CO_2 + H_2O \quad (3)$$

Next, degassed oleic acid, oleylamine, and lead halide are mixed in degassed 1-octadecene and the temperature is raised to 150–200 °C, depending on the desired size of NCs. Preheated (100 °C) Cs-ol is quickly injected and after a few seconds, the reaction is quenched in an ice-water bath (see **Figure 13**). The formed CsPbX$_3$ NCs are separated by centrifugation and dispersed in a nonpolar solvent, typically hexane. Formation of perovskite follows the salt metathesis reaction (4). However, other reaction mechanisms with various byproducts were suggested; see recently published Perspective for more details.[308]

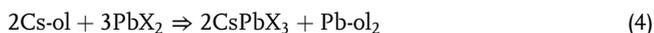

$$2Cs\text{-ol} + 3PbX_2 \Rightarrow 2CsPbX_3 + Pb\text{-ol}_2 \quad (4)$$

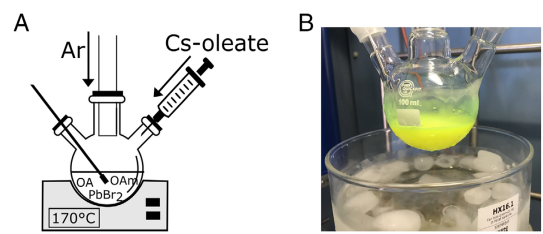

**Figure 13.** A) Schematic representation of the hot injection process. Reproduced under terms of the CC-BY license.[350] Copyright 2018, The Authors. Published by MDPI. B) Photograph of the as-synthesized CsPbBr$_3$ NCs after cooling in the ice-water bath.

Usually, the Cs-ol precipitates from 1-octadecene solution at RT and hence the preheating step before injection. A study by Lu et al.[309] points out that the reason behind this lies in strong permanent dipole of the Cs-ol molecule that significantly reduces its solubility in nonpolar solvents. To shield the polar Cs-ol molecule from the nonpolar solvent, they suggest using the ideal molar ratio 5 : 1 (oleic acid : Cs$^+$) in Cs-ol synthesis that leads to the formation of reverse micelle, Cs$^+$ ion, capped by oleate molecules. Using at least this molar ratio during the synthesis, the Cs-ol no longer precipitates at RT from the 1-octadecene solution, which allows for better reproducibility of the subsequent hot injection synthesis.

Another widely used synthesis is the so-called room-temperature precipitation (RTP) method. This method was introduced for CsPbX$_3$ NCs shortly after the introduction of the hot injection synthesis.[310] However, a similar method was first introduced for CH$_3$NH$_3$PbX$_3$ in 2015 already, by that time called ligand-assisted reprecipitation (LARP) technique.[311] So by the time of adaptation of this protocol to CsPbX$_3$ NCs, it has already been an established method for hybrid organic–inorganic perovskites preparation.[312–316]

The method is rather simple in its principle. All the precursor ions are dissolved in the polar solvent (CsX and PbX$_2$ in dimethylformamide or dimethylsulfoxide) together with oleic acid and oleylamine. This solution is then swiftly added to a nonpolar solvent in which the polar solvent is soluble (e.g., toluene) and by this step, rapid precipitation of the product is induced (5).[310]

$$Cs^+ + Pb^{2+} + 3X^- \Rightarrow CsPbX_3 \quad (5)$$

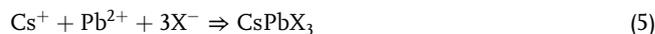

This method was also adapted for preparation of other NC shapes.[302,317–319] The example of adaptation to prepare nanoplatelets is schematically shown in **Figure 14**.

Other synthetic methods were developed for CsPbX$_3$ NCs including ultrasonic irradiation,[320–322] adaptation of classical heat-up method,[323–325] or solvothermal synthesis.[326,327]

The main idea behind many of the published alternative protocols was to enable scaling-up of the CsPbX$_3$ NC synthesis, which is not feasible for the original hot injection process. Even though hot injection yields NCs of better quality,[328,329] RTP has been identified as the viable option for the production scale-up.[330] Several studies reported successful scale-up of their protocols to a gram level.[323,331,332] However, this amount is still far from commercial production. It is also worth mentioning that some recent studies raised issues with reproducibility of the RTP synthesis; small variations of the synthetic protocol can lead to significant changes in optical properties and even compositional changes.[329,333]

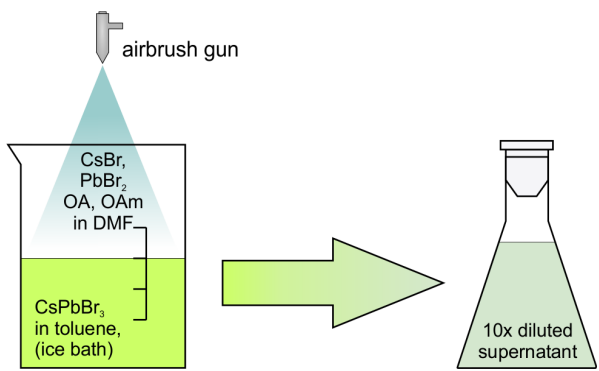

**Figure 14.** Schematic representation of RTP procedure adapted for nanoplatelets preparation using an airbrush gun. Reproduced under terms of the CC-BY license.[302] Copyright 2018, The Authors. Published by AIP Publishing.

### 10.4. Nanocomposites

The applicability of lead halide perovskites is significantly limited due to their poor chemical stability against air moisture and oxygen.[334] Recent reviews explore their stability in detail, including proposed strategies to solve this by adjusting reaction conditions, changing composition, or surface passivation.[42,289] However, for application in scintillation detectors, immobilization of the NCs in a solid monolith matrix is needed. Some recent reviews discuss progress on such matrices as well[335,336] but with different target applications in mind. Their results are readily applicable in the field of materials for scintillation detectors and the reader is encouraged to get familiar with them.

#### 10.4.1. Plastic Matrix

The easiest way to incorporate perovskite QDs in a solid matrix is to mix them with a kind of plastic. The most basic method is to dissolve a polystyrene in a toluene, mix it with $CsPbX_3$ colloidal solution in toluene, and leave it to evaporate. It has been shown that in this way, a nanocomposite can be fabricated using $CaPbBr_3$ nanoplatelets with blue emission (i.e., with strong quantum confinement effect) that, while properly stored in a cool, dark place, can maintain its blue emission.[302] However, this preliminary study showed only a weak radioluminescence response, as shown in **Figure 15**.

More recently, a similar approach was exploited to prepare $CsPbBr_3$ nanocomposite in a polymethylmethacrylate (PMMA) matrix.[337] PMMA was dissolved in $CHCl_3$ and the $CsPbBr_3$ toluene solution was added into the mixture. However, they controlled the evaporation/drying time to ≈22 h to avoid cracks and bubbles that can occur upon fast evaporation in a thick layer. They also added a perylene dyad wavelength shifter into the mixture to suppress severe reabsorption caused by a small Stokes shift that is typical for semiconductor NCs. They demonstrate radioluminescence intensities comparable with commercial $Bi_4Ge_3O_{12}$ (BGO) crystal, thanks to essentially no loss of sensitization due to perfect spectral overlap of $CsPbBr_3$ emission spectrum with perylene dyad absorption. Their nanocomposite shows good long-term stability (no degradation after 6 months of storage in air) and radiation hardness (>85% of radioluminescence intensity is retained after irradiation by 800 Gy of X-rays).

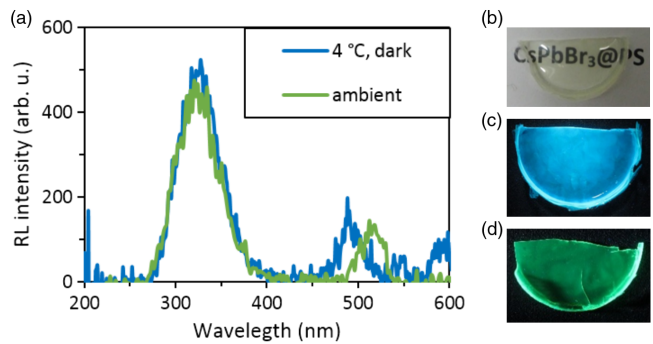

**Figure 15.** a) Radioluminescence of $CsPbBr_3$–polystyrene nanocomposite stored in a refrigerator (blue line) and under ambient conditions (green line), emission peaking at 320 nm is from polystyrene, emission around 500 nm is from $CsPbBr_3$ NCs. b) Photograph of the nanocomposite in daylight; c,d) Photographs of the nanocomposite stored in a refrigerator and under ambient conditions, respectively, under UV illumination. Reproduced under terms of the CC-BY license.[302] Copyright 2018, The Authors. Published by AIP Publishing.

Another approach is to mix the colloidal solution with a monomer and then induce polymerization. This approach is somewhat risky, because the radicals generated during the polymerization process may damage and/or strip away the surface ligands, that are already highly dynamically bound to the NC surface.[338]

This approach was exploited in its most simple and effective way by Chhangani.[339] The authors prepared $CsPbBr_3$ NCs directly in the methyl methacrylate (MMA) monomer by the modified RTP method and this solution was directly polymerized by gamma irradiation without any catalysts. However, some degree of aggregation occurred in MMA already. NCs with wide size distribution centered at 43 nm were prepared, that resulted in redshift of PL compared with other NCs prepared by the similar method. After polymerization, the emission is slightly more redshifted, and the peak is widened and less intense, suggesting some degree of degradation, but a more detailed explanation of this phenomenon was not in the scope of this study. RT scintillation response of $CsPbBr_3$ NCs toward alpha particles was demonstrated for this nanocomposite.

However, a similar nanocomposite based on polybutylmethacrylate (PBMA) was very successfully prepared by polymerization, enabled by including a ligand exchange step. In the study by Nie et al.,[340] $CsPbBr_3$ NCs were treated with BMEP (bis(2-(methacryloyloxy)ethyl) phosphate) ligand that has a methacrylate tail, that can polymerize with butyl methacrylate. This ensured high transmittance of the prepared nanocomposite (i.e., the NCs were well dispersed in the matrix). The stability was tested by immersion in water for 7 days during which no decrease in PL intensity was observed. The collected X-ray image of resistor with good spatial resolution showed a potential of $CsPbBr_3$/PBMA nanocomposites for application in X-ray imaging.

#### 10.4.2. Glass Ceramics

Incorporation of $CsPbX_3$ NCs into a glass matrix has attracted considerable attention recently, because it promises superior stability compared with the plastic matrices.[341–346] The fabrication technique is a conventional melt-quenching process. To crystallize $CsPbX_3$ NCs in the precursor glass, several hours of heat treatment are required, ranging from 340 °C for 15 h[343] to 550 °C for 4–18 h.[346]

A basic example of such nanocomposites was studied in the work by Wang et al.[341] The authors annealed borate precursor glass at 470 and 520 °C for just 1 h. They demonstrated that higher temperature led to an opaque sample due to too large NCs causing significant light scattering. On the other hand, nanocomposites treated at 470 °C had very weak radioluminescence intensity and both samples had immeasurable light yield under gamma rays. They prepared nanocomposites based on $CsPbBr_3$ and $CsPb(Cl,Br)_3$ demonstrated faster decay times with increasing Cl content, but even lower luminescence intensities. On the other hand, they demonstrated good stability of such glass ceramics. X-ray irradiation caused progressively darker spots with increasing power input (up to 12 W) with progressively lower luminescence intensity under UV illumination. However, this radiation damage was reversible by annealing at temperatures above glass transition temperature (350 °C) for 2 h, showing that this material is reusable even after high-power X-ray exposure. The authors also demonstrated good stability against heating–cooling cycle (25–300 °C).

More recent studies clearly turned their attention to improved properties by RE elements (RE) doping, for example, $Eu^{3+}$,[342,343] $Ce^{3+}$,[346] and $Lu^{3+}$.[345] Overall, RE doping caused significant improvement of transmittance that was accompanied by less light scattering in the sample, which is important to achieve high-quality X-ray imaging. RE ions in precursor glass probably serve as nucleating agents that promote rapid growth and precipitation of $CsPbX_3$, thus preparing more uniformly distributed, smaller crystals.[342,344] A similar effect was achieved by introducing AgCl into the precursor glass.[343]

Also, the radioluminescence performance has been significantly improved by RE doping. Not only the emission from $CsPbX_3$ was enhanced,[346] but in case of $Eu^{3+}$ doping, strong emission from dopant emerged to accompany the enhanced $CsPbX_3$ emission, that caused larger Stokes shift in such a material, effectively suppressing self-absorption.[342,344]

Improved stability against various conditions was successfully tested, for example, against humidity,[342,344,346] heat–cooling cycles,[343] and X-ray exposure.[344] Exposing the material to high radiation doses (for 120 h by dose rate of 8 $mGy_{air}\,s^{-1}$) caused the radiation damage, but it was reversible by thermal annealing at 300 °C for 3 h.[345] Overall, this glass ceramics based on $CsPbX_3$ NCs was proven very promising for scintillation application. High-quality X-ray images were achieved[342,344,345] with spatial resolution as high as ≈15–17 lp $mm^{-1}$.[342,345]

To further improve transmittance and radioluminescent properties, the use of tellurite glass was proposed.[343] Tellurite glass is beneficial thanks to its larger density (>5.0 g $cm^{-3}$) that results in better stopping power. Moreover, its refractive index is large (>2.0), leading to smaller mismatch with $CsPbX_3$ NCs, thus significantly improving transparency of the resulting nanocomposite.

#### 10.4.3. Polycrystalline Thin Films

Lead halide perovskite NCs have been selected as potential candidates for ultrafast detector development for TOF-PET and high-energy physics.[347] The proposed strategy is to fabricate a sandwich-like structure combining layers of a single-crystal scintillator and semiconductor NCs, as shown in **Figure 16**. The bulk scintillator (e. g. LYSO:Ce, BGO, or $Gd_3(Ga,Al)_5O_{12}$:Ce–GGAG:Ce) possesses high stopping power and energy resolution, while NCs (e. g. $CsPbX_3$ or CdSe) have ultrafast decay times. A combination of such materials results in improved coincidence time resolution (CTR) that ultimately leads to better spatial resolution of TOF-PET detectors.[347] This concept was already tested for CdSe nanoplatelets and LYSO:Ce pixel.[348]

The first preliminary study following this concept was fabricating $CsPbBr_3$ thin films on LYSO:Ce bulk scintillator wafer.[349] $CsPbBr_3$ NCs were prepared by RTP method and then drop cast on the LYSO:Ce wafer. Enhanced $CsPbBr_3$ radioluminescence was observed in the steady-state radioluminescence spectrum and fast decay components of $CsPbBr_3$ NCs were preserved in both photo- and radioluminescence decays. However, the scintillation decay measurement had rather a poor dynamic range, suggesting poor scintillation light output of such samples. The reason for such performance probably lies in significant contamination by the $Cs_4PbBr_6$ phase that was later found to be detrimental for the scintillation light output of $CsPbBr_3$ NCs.[329]

Recently, $CsPbBr_3$ thin films were prepared on GGAG:Ce scintillating wafer.[350] The NCs prepared by the hot injection method featured pure $CsPbBr_3$ phase (no presence of $Cs_4PbBr_6$ phase) and thin films were fabricated using a spin-coating method. A synergic effect was demonstrated; both $CsPbBr_3$ and GGAG:Ce radioluminescence emissions were enhanced in the steady-state radioluminescence spectra, the former probably by absorption and subsequent reemission of GGAG:Ce emission by $CsPbBr_3$ layer and the latter by light-guiding effect of cracks in the $CsPbBr_3$ layer. This enhancement was also observed in the subnanosecond time gate, suggesting the applicability of such composites in future fast timing detectors.

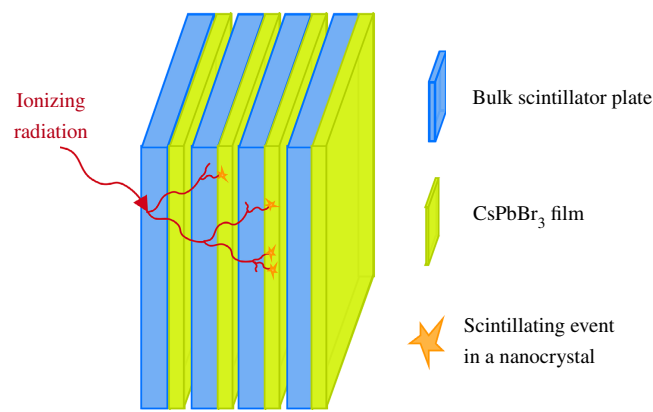

**Figure 16.** Schematic representation of the proposed sandwich pixel for ultrafast detector system.

However, such unprotected thin films suffer from poor stability. To improve thin-film stability, solution-protected annealing strategy was successfully implemented.[351] After purification of hot injection-prepared NCs, the colloidal solution returned to the mixture of 1-octadecene, oleic acid, and oleylamine and NCs were treated at 80 °C. Obtained NCs had better crystallinity, a well-passivated surface, and showed improved luminescence properties and stability toward the heating–cooling cycle, air, UV, electron, and X-Ray irradiation. X-ray imaging capability of a thin film fabricated using such treated NCs was also demonstrated.

Similarly, as for bulk monoliths of glass or plastic, encapsulation in matrices has been explored for enhancing stability of thin films as well. Recently, $CsPbBr_3$ NCs embedded in $Al_2O_3$ arrays have been shown to have significantly improved stability compared with conventional dense nanocrystalline thin films and X-ray imaging capability with superior resolution of 250 lp $mm^{-1}$.[352] The pore diameter was 300 nm, with 20 μm pore depth. $CsPbBr_3$ NCs were prepared beforehand by the hot injection method and then incorporated into $Al_2O_3$ arrays by negative pressure filling, forming a micropillar structure. These micropillars serve as optical waveguides that significantly improve spatial resolution compared with the standard thin film of the same thickness (20 μm).

Another approach was explored by Zhao et al.,[353] who fabricated perovskite-based flexible detectors by incorporating $MAPb(I_{0.9}Cl_{0.1})_3$ into a 100 μm-thick porous nylon membrane. They prepared ultraconcentrated saturated $MAPbI_3$ solutions (2.2 mol $dm^{-3}$) in a mixture of 2-methoxyethanol and methylammonium chloride. This solution was repeatedly infiltrated into the membrane pores by negative pressure filling and subsequent annealing at 150 °C for 2 h. This led to a fully filled membrane by interconnected perovskite crystals allowing efficient charge transport and providing high stopping power of the composite. Flexible scintillation screen with resolution of 3.5 lp $mm^{-1}$ was achieved and its applicability in industrial X-ray imaging and defect detection was demonstrated.

## 11. Conclusion

We gave an extensive overview of the R&D of halide scintillators in the last decade going from the bulk single crystals through microstructured systems to scintillating NCs. In bulk single crystals, the multicomponent compositions are explored to achieve optimal performance and create an application-specific scintillator. Moreover, codoping is used to improve the performance of both classical and recent innovative scintillators. Ceramic technology has been also applied for several material compositions to create transparent monolites. In microstructured scintillators, light guiding is exploited to significantly improve 2D resolution in X-ray imaging techniques. Complex eutectic compositions are explored to maximize the difference in refraction index and to enable the detection of both γ (X-ray) and neutrons. In lead halide perovskite NCs, the synthesis routes are being optimized and NC embedding in various matrices is used to increase their stability. Moreover, ensembles of NC-based materials with classical scintillators (so-called metascintillators) possess both high stopping power and light output of classical scintillator and ultrafast emission of NCs. Research in all the above-mentioned fields achieved significant progress in the last decade and several materials emerged as candidates for commercialization. With new applications requiring high-performance scintillators with tailored parameters, modern halide scintillators could play an increasing role in the future scintillation market.

## Supporting Information

Supporting Information is available from the Wiley Online Library or from the author.

## Acknowledgements

The authors would like to dedicate this article to a colleague and a long-term senior researcher of the Institute of Physics of the Czech Academy of Sciences Dr. Karel Nitsch. He significantly contributed to the research of dielectric materials for optical applications (based on e.g., halides, oxides, and phosphate glasses), development of crystal growth technology, and education and popularization of the field in the crystal growth community. He was a leading coordinator and founder of the Czechoslovak Association for Crystal Growth in 1990 and served as its chairman until 2012. Further, he periodically organized domestic conferences and seminars and contributed into international journals and periodicals. Financial support of the Czech Science Foundation, grant number GA20-06374S, is acknowledged.

## Conflict of Interest

The authors declare no conflict of interest.

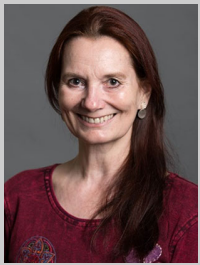

**Eva Mihóková** is a senior scientist at FZU Institute of Physics, Czech Academy of Sciences, and also an associate professor at the Czech Technical University in Prague. She obtained her Ph.D. (1996) in the solid-state physics in the Institute of Physics, Academy of Sciences of the Czech Republic. Her research includes study of decay kinetics of heavy $ns^2$ ions and rare earth ions in oxides and halides, luminescence and scintillation mechanisms, quantum confinement effects in the foreign aggregated phases in selected halide matrices, development of nanocomposites for fast timing detection systems, and biomedical applications.

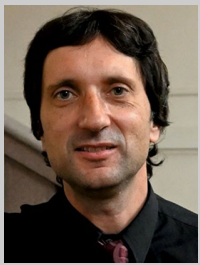

**Václav Čuba** is an associate professor at the Czech Technical University in Prague. He obtained his Ph.D. (2003) in nuclear chemistry at the Faculty of Nuclear Sciences and Physical Engineering. He has been focused on research in applied radiation chemistry and processing, photochemistry, materials chemistry, and nanotechnology and he leads a subgroup focused on synthesis of scintillating NCs and nanomaterials for a wide array of applications.

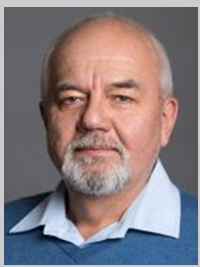

**Martin Nikl** graduated in 1981 from the Faculty of Nuclear Science and Physical Engineering, Czech Technical University, and obtained his Ph.D. in 1986 at the Institute of Physics, CAS. He became an associated and full professor at Czech Technical University in 2015 and 2021, respectively. His research interests include luminescence and scintillation mechanisms in wide-bandgap solids, energy-transfer processes, and the role of material defects in them. He currently serves as the department head at the Institute of Physics, Academy of Sciences of the Czech Republic.